\documentclass[12pt]{iopart}
\usepackage{graphicx}
\usepackage{dcolumn}
\usepackage{bm}
\usepackage{amssymb}
\begin{document}
\input psfig.tex
\def\be{\begin{equation}}
\def\ee{\end{equation}}
\def\ba{\begin{eqnarray}}
\def\ea{\end{eqnarray}}
\newcommand{\s}{\sigma}
\title{Hartree-Fock energy of a density wave
in a spin polarized two-dimensional electron gas}

\author{Juana Moreno and D. C. Marinescu}
\address{Department of Physics, Clemson University, Clemson, SC 29634}
\date{\today}
\begin{abstract}

We calculate the Hartree-Fock energy of a density-wave in a spin
polarized two-dimensional electron gas using a short-range
repulsive interaction. We find that the stable ground state for a
short-range potential is {\it always}  either the 
paramagnetic state or the uniform ferromagnetic state. The energy
of a density-wave state 
is, however, reduced by a factor proportional
to $(1 -\zeta^2)$, where $\zeta$ is the polarization of
the electron gas. Since this situation occurs in the most
unfavorable conditions (short-range repulsive interaction) it is
therefore conceivable that by including higher order many-body
corrections to the interaction a density-wave ground state is
indeed found to be stable.

\end{abstract}

\maketitle

The possible existence of a non-uniform ground state for a
two-dimensional (2D) electron system in the presence of a magnetic
field has been proposed in explaining the results of a
number of recent experiments. Several examples of such puzzling
data are the unidirectional charge-density-wave states that appear
in partially filled Landau levels \cite{lilly99,du99}, the
incompressible, inhomogeneous insulating phase in p-GaAs/AlGaAs
heterostructures displaying metal-insulator transition
\cite{Dultz00,Ilani00}, and the unusual magneto-optical properties
of the ferromagnetic phase of p-type $Cd_{1-x}Mn_xTe$ quantum
wells \cite{Kossacki}.

Numerical calculations\cite{Attacalite} have definitely proved that
at low densities an unpolarized electron gas forms a Wigner crystal. 
However, a study of point defects in the two-dimensional
Wigner crystal suggests that 
the quantum melting could be continuous rather than first 
order\cite{Candido01}, leaving open the possibility
of inhomogeneous intermediate phases.

The idea of a non-uniform ground state in a paramagnetic 2D
electron system has surfaced before. Prior to the discovery of the
quantum Hall effect, it was argued that in 2D GaAs-type
structures the ground state was a charge-density-wave
\cite{Fuku79}. More recent experimental\cite{lilly99,du99} and
theoretical\cite{koulakov96,moessner96} results point indeed to
the existence of charge-density-wave states in partially filled
higher Landau levels on account of the quasi-one-dimensional
electron motion.

In this work we are interested in the possible formation of density waves in
a 2D electron system that exhibits a very large Zeeman splitting.
This situation occurs in II-VI dilute-magnetic semiconductor structures
where the effective Land\'{e} factor is up to thousands of times the 
band value. This is exactly opposite to the case in GaAs, where the
cyclotron energy is dominant. For example, in GaAs the ratio of the cyclotron
frequency to the Zeeman splitting is around 14, while the same ratio is
close to 1/20 in $Zn_{1-x}Cd_xSe$.

Quantum Monte Carlo results show
that the exchange-correlation hole is larger in the polarized electron
gas\cite{Tan89,Rapisarda96} suggesting that a highly polarized system
is prone to density instabilities. Whether or not the instability
develops before the Wigner crystallization and its dependence on
the degree of polarization has not been studied yet. Here, we
calculate the Hartree-Fock (HF) energy of a spin density wave
state in a spin polarized system
for a delta-function repulsive interaction, the most unfavorable situation
for the development of a density-wave instability in an isotropic system.

It is well known that in three dimensions, within the HF
approximation and for an unscreened Coulomb interaction, the
paramagnetic state is unstable with respect to formation of a spin
density wave with wave vector near $2 k_F$ \cite{Over62}.
However, for a short-range potential the stable HF solution is
{\it always} either the usual paramagnetic state or the uniform
ferromagnetic state \cite{Herring}. We find that this result holds
also in the 2D polarized system. In this case, however, the
difference in energy between a density wave state with momentum
$\bf q$, $E({\bf q})/N$, and the ferromagnetic state, $E({\bf
q}=0))/N$, is reduced by a factor proportional to $(1 -\zeta^2)$,
where $\zeta=(n_{\uparrow}-n_{\downarrow})/n$ is the polarization
of the electron gas, 
\be 
\frac{E({\bf q})-E({\bf q}=0)}{N}=
{\bf q}^2 \frac{1 -\zeta^2}{4}. 
\ee  
If the background charge is allowed to relax, as in the 
deformable jellium model, our result applies to both
spin- and charge-density waves \cite{Over68}. 

Since the effective electronic interaction in real systems is somewhere in
between the short-range  delta-function and the unscreened Coulomb
potential our result suggest the possibility of an inhomogeneous
density wave ground state in highly polarized two-dimensional
electron systems. If  this inhomogeneous state exists it will be
most probably a charge density wave, since electronic correlations
in real systems favor charge-density over spin-density wave
instabilities \cite{Over68}. Below we present the details of our
analysis.

A density wave develops when the correlation function between an
electron with momentum {\bf k} and spin $\sigma$ and another
electron with momentum {\bf k+q} and spin $\sigma'$, $\langle
\psi^{\dagger}_{{\bf k} \sigma} \psi_{{\bf k+q} \sigma'}\rangle$
becomes finite. When $\sigma = \sigma'$, a charge-density wave is
formed, while $\sigma \neq \sigma'$ corresponds to a spin-density
wave. The interacting electron system Hamiltonian is
diagonalized by a canonical transformation that introduces a new
set of operators:
 \ba \psi_{\bf k}^{lower}=\cos{(\theta_{\bf k}/2)}
\psi_{\bf k-\frac{q}{2} \sigma}+
\sin{(\theta_{\bf k}/2)} \psi_{\bf k+\frac{q}{2} \sigma'} \\
\psi_{\bf k}^{upper} =- \sin{(\theta_{\bf k}/2)} \psi_{\bf
k-\frac{q}{2}\sigma}+ \cos{(\theta_{\bf k}/2)} \psi_{\bf
k+\frac{q}{2} \sigma'}, \ea where $\theta_{\bf k}$ is the coupling
parameter and $\psi_{\bf k}^{lower} (\psi_{\bf k}^{upper})$ refers
to the new lower (upper) band excitations.

The Hartree-Fock ground state energy
is a function of {\bf q} and the parameters $\theta_{\bf k}$ \cite{Herring}.
For a system with $N$ electrons, volume $V$ and polarization $\zeta$ the total
ground state energy is:
\ba
E({\bf q})=\frac{1}{2m}\sum_{\bf k} \left[ {\bf k}^2 +\frac{\bf q^2}{4} -
({\bf k} \cdot {\bf q}) \cos{\theta_{\bf k}} \right]
-\frac{H}{2}\left[ \sum_{\bf k} \cos{\theta_{\bf k}}
- N \zeta \right]\nonumber\\
-\frac{1}{2 V} \sum_{\bf k k^{'}}  v({\bf k-k^{'}})
\cos^{2}{[(\theta_{\bf k}-\theta_{\bf k'})/2]} \ea where the first
term is the kinetic energy, the second is the effective Zeeman
energy\footnote{The spin magnetization is written as
$\displaystyle M=\frac{N \zeta}{2}=\frac{1}{2} \sum_k
\cos{\theta_{\bf k}}$.}, needed to fulfill the constraint of
constant  polarization,
 and the last one
the Coulomb and exchange energies.

For a delta-function interaction with strength $ v({\bf k-k^{'}})=v_0$
the total ground state energy  in atomic units
\footnote{${\bf k}$ and ${\bf q}$ are in units of inverse
Bohr radius ($a^{-1}_0$) and the energy in rydbergs.} becomes:
\ba
E({\bf q})=\sum_{\bf k} \left[ {\bf k}^2 +\frac{{\bf q}^2}{4} -
({\bf k} \cdot {\bf q}) \cos{\theta_{\bf k}} \right]
-\frac{H}{2}\left[ \sum_{\bf k} \cos{\theta_{\bf k}} - N \zeta \right]\nonumber\\
-\frac{v_0}{4 V} \left[ \left( \sum_{\bf k} \cos{\theta_{\bf k}}\right)^2
+\left( \sum_{\bf k} \sin{\theta_{\bf k}}\right)^2 -N^2 \right].
\label{eq:tot.energy}
\ea

$E({\bf q})$ reaches a minimum with respect to $\theta_{\bf k}$
when \ba \tan{\theta_{\bf k}}=\frac{\kappa(v_0,\zeta) }{2 {\bf
k}\cdot {\bf q} + H^{'}(v_0,\zeta)} \hspace{1cm}{\rm and}
\hspace{1cm} \kappa(v_0,\zeta)=\frac{v_0}{V} \sum_{k}
\sin{\theta_k} \label{eq:minimum} \ea

where $n=N/V$, $\displaystyle \zeta=\frac{1}{N} \sum_{k} \cos{\theta_k}$ and
$H'=H+ v_0 \zeta n$.

The quasi-particle energy in the ground state when only the lower
band is occupied is given by: \ba \epsilon^{lower}_k = {\bf k}^2 +
\frac{{\bf q}^2}{4} + \frac{n v_0}{2} -\frac{1}{2} \sqrt{[2 {\bf
k}\cdot {\bf q} + H^{'}(v_0,\zeta)]^2 +
  \kappa^{2} (v_0,\zeta)}.
\ea

The solution for ${\bf q'}= \lambda {\bf q}, v'_0=v_0/\lambda,
V'=V/\lambda^3$ is homologous to that for
${\bf q}, v_0, V$, and results in $\kappa'=  \lambda^2 \kappa$
and $H'=  \lambda^2 H$. Then, from a numerical point of view,
the simpler way to minimize the total energy is to take
$\kappa(v_0,\zeta)=1$ and a fixed value for the parameter
$H'$, select the  maximum occupied energy in the lower band,
and then compute $n, \zeta, v_0$ and $E({\bf q})$ for this choice
of parameters using Eqs.~(\ref{eq:tot.energy}) and (\ref{eq:minimum}).

\begin{figure}
\begin{minipage}{\linewidth}
\centerline{\includegraphics[width=0.65\textwidth]{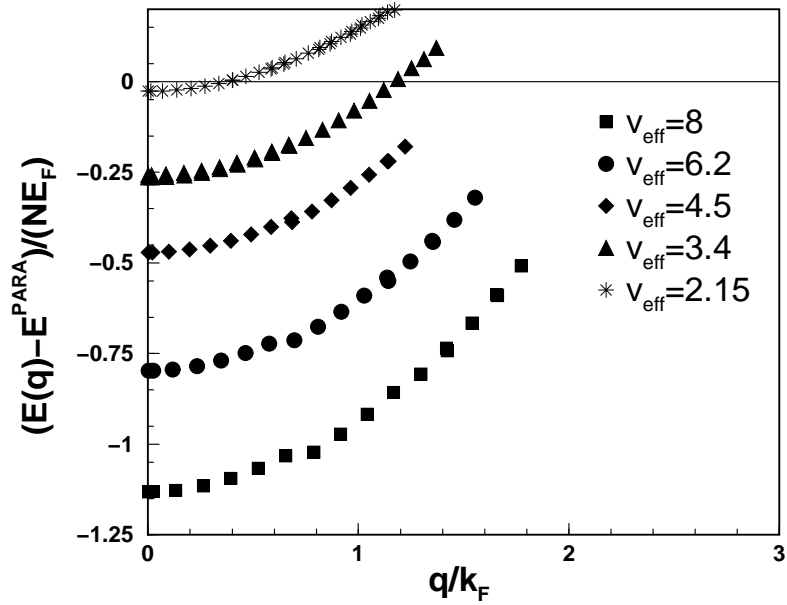}}
\end{minipage}
\caption {Normalized Hartree-Fock energy of the spin density wave state,
$(E({\bf q})-E^{PARA})/(N E_F)$ versus its normalized momentum,
${\bf q}/k_F$,   for  $\zeta =0.5$ and different values
of the effective potential strength, $v_{eff}$, as indicated in the legend.}
\label{fig:sdweng.pol50}
\end{figure}

Fig.~\ref{fig:sdweng.pol50} displays our results for the
normalized energy per particle of the spin density wave state,
$E({\bf q})/(N E_F)$, as function of normalized
momentum, ${\bf q}/k_F$, for $\zeta=0.5$ and several values of the effective
potential strength, $\displaystyle v_{eff}=\frac{n v_0}{E_F}= \frac{v_0}{2 \pi}$.
The energy is measured with respect to the
energy of the paramagnetic state with the same
density and polarization,
$\displaystyle \frac{E^{PARA}}{N E_F}=\frac{1}{2} (1 +\zeta^2)
+\frac{v_{eff}}{4 } (1 -\zeta^2)$.
For values of $v_{eff} >2$ there is a range of momenta where the density
wave states are lowest in energy than the non-magnetic state.
However, {\it always}
the absolute minimum is at ${\bf q}=0$, the ferromagnetic state.
The range of momenta where the density
wave states are stable increases with the strength of the interactions,
while the energies get reduced.

\begin{figure}
\begin{minipage}{\linewidth}
\centerline{\includegraphics[width=0.65\textwidth]{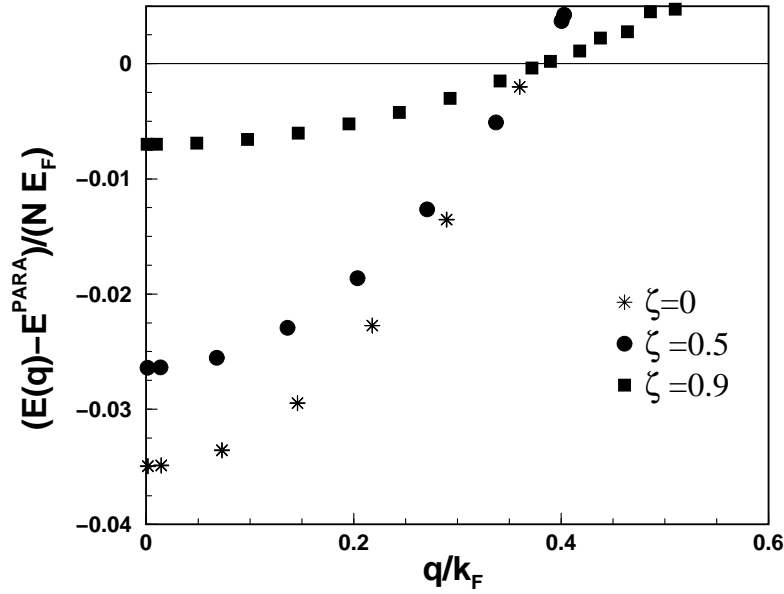}}
\end{minipage}
\caption {Normalized Hartree-Fock energy of the spin density wave state,
$(E({\bf q})-E^{PARA})/(N E_F)$ versus its normalized momentum,
${\bf q}/k_F$,   for  an effective interaction strength of
$v_{eff}=2.15$ and $\zeta=0$ (stars),  $\zeta=0.5$ (circles)
and $\zeta=1$ (squares).}
\label{fig:sdweng.v215}
\end{figure}

Fig.~\ref{fig:sdweng.v215} shows the normalized energy for
$v_{eff}=2.15$ and three values of the polarization, $\zeta=0,0.5,0.9$.
The range of momenta where the density
wave state is lower in energy is the same for all values of polarization.
However, the energies increase with  polarization.

\begin{figure}
\begin{minipage}{\linewidth}
\centerline{\includegraphics[width=.65\textwidth]{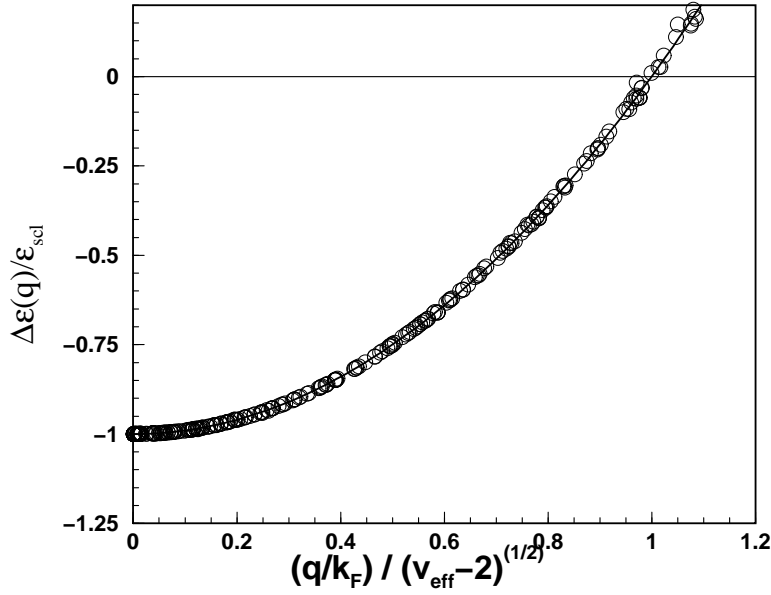}}
\end{minipage}
\caption {Scaled Hartree-Fock energy per particle of the
spin density wave state, $\Delta \epsilon({\bf q})/\epsilon_{scl}$,
as a function of the scaled momentum, $\tilde{q}=q/(k_F \sqrt{v_{eff}-2})$.
Circles correspond to different values of
$v_{eff}$ (between 2 and 8) and polarization $\zeta=0,0.5$ and $0.9$.
The solid line corresponds to the scaling relation Eq.~(\ref{eq:scaling}).}
\label{fig:scaling}
\end{figure}

By scaling the energies and momenta we find that all our numerical
results overlap on a single curve.
Fig.~(\ref{fig:scaling}) displays our results for
the energy per particle relative to the energy of the
paramagnetic state
$\displaystyle \Delta \epsilon({\bf q})=
(E({\bf q})-E^{PARA})/(N E_F)$ divided over
$\epsilon_{scl}= (v_{eff}-2)(1-\zeta^2)/4$ as a function of
the scaled momentum
$\displaystyle \tilde{q}=\frac {q}{k_F \sqrt{v_{eff}-2}}$.
We have studied a full range of values of $v_0$ and $\zeta$.
All the data are in excellent agreement with
the surprising simple scaling relation:
\be
\frac{\Delta \epsilon({\bf q})}{\epsilon_{scl}}=
\tilde{q}^2 -1.
\label{eq:scaling}
\ee
Therefore, the HF energy
of a two-dimensional density wave (Eq.~(\ref{eq:tot.energy}))
is a quadratic function of the wave-vector
for any value of the polarization and strength interaction.

In conclusion, we have found that independently of the
polarization, for $\displaystyle v_{eff}> 2$ and wave-vector
$\displaystyle {\bf q}< k_F \sqrt{v_{eff} -2}$ the density wave
ground state is lower in energy than the paramagnetic state. In
all cases, the absolute minima is the ferromagnetic state with
${\bf q}=0$, but the depth of the minima scales with the
polarization as $(1-\zeta^2)$. As a consequence, in a highly
polarized electron gas the density wave states are very close in
energy to the ferromagnetic instability. 
Because a short range potential is the most unfavorable  for the
development of density instabilities, that fact suggests that
in real systems the ground state might be a  density wave state in some
range of densities and spin polarizations \cite{Moreno03}.
Since in two dimensions
transitions to long-range order phases take place at zero
temperature, the extension of our arguments to the geometry of
real systems, quasi-two-dimensional heterostructures, thus
allowing direct comparison with experimental results
\cite{Kossacki}, is currently being investigated.

 {\bf Acknowledgments}

We are grateful to T. Dietl, R.~S. Fishman, M.~R. Geller
and J.~R. Manson for enlightened discussions.
We acknowledge the financial support provided by the
Department of Energy, grant no. DE-FG02-01ER45897.

\end{document}